 \documentclass[manuscript]{acmart}
\AtBeginDocument{%
  \providecommand\BibTeX{{%
    \normalfont B\kern-0.5em{\scshape i\kern-0.25em b}\kern-0.8em\TeX}}}

\setcopyright{acmcopyright}
\copyrightyear{2018}
\acmYear{2018}
\acmDOI{XXXXXXX.XXXXXXX}

\acmConference[Conference acronym 'XX]{Make sure to enter the correct
  conference title from your rights confirmation emai}{June 03--05,
  2018}{Woodstock, NY}
\acmPrice{15.00}
\acmISBN{978-1-4503-XXXX-X/18/06}

\begin{document}

\title{User Archetypes and Information Dynamics on Telegram: COVID-19 and Climate Change Discourse in Singapore}


\author{Val Alvern Cueco Ligo}
\affiliation{%
 \institution{Nimblemind} 
\country{USA}
}

\author{Lam	Yin Cheung}
\affiliation{%
 \institution{Nimblemind} 
 \country{USA}
 }


 \author{Roy	Ka-Wei Lee}
\affiliation{%
\institution{Singapore University of Technology and Design}
 \country{Singapore}
 }

 \author{Koustuv	Saha}
\affiliation{
 \institution{University of Illinois Urbana-Champaign}
 \country{USA}
 }
 
\author{Edson C. Tandoc Jr.}
\affiliation{
 \institution{Nanyang Technological University}
 \country{Singapore}
 }

 \author{Navin Kumar}
\affiliation{%
 \institution{Nimblemind}
  \country{USA}
}

\renewcommand{\shortauthors}{Kumar et al.}
\begin{abstract}
Social media platforms, particularly Telegram, play a pivotal role in shaping public perceptions and opinions on global and national issues. Unlike traditional news media, Telegram allows for the proliferation of user-generated content with minimal oversight, making it a significant venue for the spread of controversial and misinformative content. During the COVID-19 pandemic, Telegram's popularity surged in Singapore, a country with one of the highest rates of social media use globally. We leverage Singapore-based Telegram data to analyze information flows within groups focused on COVID-19 and climate change. Using k-means clustering, we identified distinct user archetypes, including Strategic Disruptor, Empirical Enthusiast, Inquisitive Moderate, and Critical Examiner, each contributing uniquely to the discourse. We developed a model to classify users into these clusters (Precision: Climate change: 0.99; COVID-19: 0.95). 
\end{abstract}




\maketitle

\section{Introduction}
Social media platforms play an increasingly central role in shaping public perceptions and opinions on global and national issues \cite{sun2024social}. Unlike traditional news media, which provides a curated flow of information, social media allows for user-generated content to proliferate, often with minimal oversight \cite{tu2022viral}. This characteristic makes platforms like Telegram particularly influential in forming public opinion on contentious topics \cite{schulze2022far}. Telegram is a free cloud-based instant messaging platform \cite{urman2022they}. Telegram offers several options for engagement, including private one-on-one conversations, group chats, and both private and public channels controlled by administrators. Telegram does not partake in extensive content-moderation policies compared to apps like Facebook, Instagram, and Twitter \cite{zhong2024proud}. As a  result, Telegram tends to have significant controversial and misinformative content. 

During the COVID-19 pandemic, Telegram saw a significant rise in popularity in Singapore, becoming a major venue for discussions about the pandemic and other pressing issues such as climate change \cite{ng2020analyzing}. Analyzing Singapore-based social media data is of great importance as Singapore has one of the highest rates of social media use globally, making it a rich data source for studying online information flows \cite{tandoc2021developing}. Moreover, the Singapore government plays a proactive role in managing information dissemination, and thus better understanding of Telegram discussions in Singapore may offer insights into effective misinformation management globally \cite{abdou2021good,hien2023fatherhood}. As described, Telegram’s closed messaging system and relatively lax moderation policies create a fertile ground for the spread of misinformation. This phenomenon was evident during the COVID-19 pandemic, where misinformation regarding the causes, remedies, and policies related to the virus was rampant in Singapore-centric Telegram groups \cite{ng2020analyzing}. The spread of such misinformation has significant implications, potentially undermining public health efforts and contributing to social discord \cite{chen2022us,nainani2022categorizing}. The dynamics within these groups often reflect broader societal attitudes and behaviors, with some individuals actively disseminating misinformation, while many others remain passive consumers, undecided about the accuracy of the information they encounter. Understanding the flow of information within these closed systems is critical for developing strategies to mitigate misinformation. Information pathways refer to the routes through which information travels across different users and groups within a platform. Historically, disinformation campaigns, such as those concerning AIDS in the 1980s and Ebola in 2014, have highlighted the importance of mapping these pathways to preempt and counteract the spread of false information \cite{kramer2020lessons,sell2020misinformation}. These pathways often involve a few key individuals who set the agenda, followed by others who propagate the information, whether accurately or not \cite{lotan2011arab}.

We seek to identify user archetypes within Telegram groups on COVID-19 and climate change, and examine how these users contribute to the dissemination of information. By COVID-19, we refer to any discussion group which relates to COVID-19. COVID-19 and climate change will be treated as independent topics throughout the paper. COVID-19 was selected as a topic given the rapid spread of COVID-19 misinformation in Singapore \cite{ei2023understanding}. We selected climate change as Singapore, a low-lying island, is directly influenced by the rising sea level caused by climate change \cite{yang2020climate}. By analyzing user roles and information flow, we aim to uncover patterns that can inform effective strategies for combating misinformation. We aim to provide a nuanced understanding of how digital interactions on closed messaging systems shape perceptions around national issues in Singapore. We propose the following research question: What are the archetypes of users in Singapore-based Telegram groups around COVID-19 and climate change? As an exploratory analysis, we seek to develop a model to classify individuals into archetypes based on the preliminary clusters detailed in the research question. 


\section{Related Work}
\textbf{Information flow online}
Understanding how information flows in online environments is essential for managing misinformation effectively. Gomez Rodriguez et al. (2013) explored the dynamics of information pathways in online media, highlighting how information spreads across networks \cite{gomez2013structure}. They found that while information pathways for recurring topics are stable, those for ongoing news events are more volatile, with clusters of media sites emerging and vanishing rapidly. Similarly, Stewart et al. (2019) introduced the concept of information gerrymandering, where network structures can distort collective decision-making \cite{stewart2019information}. Their analysis demonstrated that strategically placed zealots within a network could bias outcomes, highlighting how the design of social networks can influence information flow and public opinion. Barriers to effective online collaboration were examined by Smithson et al. (2012), who identified expectations of interaction, technological unfamiliarity, and differing academic norms as significant hindrances \cite{smithson2012online}. They suggested focused forums and technical support to overcome these barriers, emphasizing the challenges of fostering productive online interactions. This is particularly relevant when considering the findings of Singh et al. (2020), who analyzed COVID-19 discussions on Twitter and found a significant relationship between information flow and new COVID-19 cases \cite{singh2020first}. They noted the presence of myths and poor-quality information, though these were less dominant than other crisis-specific themes, underscoring the role of social media in pandemic information dissemination. The influence of fake news on political processes was highlighted by Bovet and Makse (2019), who analyzed Twitter activity during the 2016 US presidential election \cite{bovet2019influence}. They identified networks of influential spreaders of fake news and traditional news, illustrating the dynamics of misinformation dissemination. Our study seeks to fill the gap by analyzing how user archetypes are key to understanding information pathways within closed messaging systems. 

\textbf{Controversial information on Telegram}
Telegram's design prioritizes user security and minimal content moderation, making it appealing for communities banned from mainstream platforms, such as conspiracy content creators and far-right movements\cite{wischerath2024spreading}. This has led to the proliferation of controversial and often harmful content on the platform. Several studies have highlighted how deplatforming from mainstream social media can push users to alternative platforms like Telegram, where they continue to propagate their ideologies. Bryanov et al. (2021) examined the migration of right-wing users to Telegram following Donald Trump’s ban from major social media platforms, noting a significant increase in user base and activity in right-wing communities \cite{bryanov2021other}. Similarly, Zhong et al. (2024) focused on the Proud Boys' presence on Telegram, documenting their growth and interaction with other far-right groups, which was catalyzed by deplatforming actions and significant events like the January 6th U.S. Capitol attack \cite{zhong2024proud}. Telegram's affordances also enable the proliferation of scams, fakes, and conspiracy movements \cite{la2021uncovering}. Their large-scale analysis revealed the presence of illegal activities and the spread of misinformation through fake and clone channels. They highlighted the challenges of detecting and managing deceptive content on Telegram, even with automated tools. Furthermore, Telegram facilitates the spread of hate speech and propaganda. Sulzhytski (2022) analyzed pro-government Telegram channels in Belarus, identifying the use of aggressive hate speech to demonize government opponents \cite{sulzhytski2022opposition}. The spread of misinformation and conspiracy theories is not limited to political contexts. Schlette et al. (2023) studied Dutch anti-vaccination communities on Telegram, finding that messages often contained shared identity elements and misinformation, with a significant portion promoting negative emotions \cite{schlette2023information}. This aligns with findings from Farrell-Molloy (2022), who explored the eco-fascist subculture on Telegram, identifying dominant narratives that blend environmentalism with far-right ideologies \cite{farrell2022blood}. These studies collectively emphasize the complex ecosystem of controversial information on Telegram. They show how deplatforming from mainstream social media can push users to Telegram, where minimal moderation allows misinformation and extremist ideologies to thrive. Our study aims to further understand these dynamics by analyzing user roles and information flow within Singapore-based Telegram groups discussing COVID-19 and climate change.

\textbf{Telegram use in Singapore}
Telegram has played a crucial role in information dissemination and public discourse in Singapore, particularly during the COVID-19 pandemic. Several studies have explored this phenomenon, highlighting the platform's impact on public opinion, misinformation, and community engagement. Ng and Loke (2020) analyzed a large Singapore-based COVID-19 Telegram group with over 10,000 participants \cite{ng2020analyzing}. They found that user participation peaked following significant public health announcements, such as the Ministry of Health raising the disease alert level. However, this heightened engagement was not sustained over time. The study also highlighted the prevalence of criticism around government-identified misinformation, indicating a critical approach by the participants towards the information presented on the platform. In a complementary study, Chen and Neo (2020) examined the content of messages in several COVID-19-related Telegram groups using topic modeling \cite{chen2020topics}. They identified three main areas of concern among the public: buying and selling protective equipment, information about masks, and signs and symptoms of the disease. Abidin (2020) conducted an ethnographic study on meme factories in Singapore and Malaysia \cite{abidin2020meme}. This study explored how these groups adapted their content and strategies in response to the pandemic. Meme factories, which are coordinated networks of content creators, played a significant role in both spreading and countering misinformation. They utilized various vernaculars to provide public service messaging and social critique, demonstrating the diverse ways in which different actors on Telegram influence public opinion. These studies collectively illustrate the multifaceted role of Telegram in Singapore during the COVID-19 pandemic. They show that Telegram serves not only as a platform for information dissemination but also as a space for critical engagement, commerce, and socio-political discourse. We seek to add to this work and develop user archetypes to better understand the information flow.

\section{Methods}

\begin{table}[!htbp] \centering 
\begin{tabular}{ |p{5cm}|p{5cm}| }
 \hline
 Dataset & No. of Words\\
 \hline 
SG Covid-19 Recovery Support Group & 540,412\\
 \hline 
LifeOfUnvaxxed Discussion & 357,450\\
 \hline 
SG Corona Freedom Lounge & 2,334,981\\
 \hline 
SG Defense against Vaxx & 148,829\\
 \hline 
Healing the Divide Discussion & 4,277,914\\
 \hline 
CovidSurvivors & 577,351\\
 \hline  
SG The Other Side Of Climate Change & 21,599,409 \\
 \hline 
\end{tabular}
\caption{Telegram Groups}
\label{tab:table_1}
\end{table}

\textbf{Data collection and preparation}
We first selected three content experts who had published at least ten peer-reviewed articles in the last three years around COVID-19, and similarly chose another three experts on climate change. We ensured the content experts conducted research on COVID-19/climate change in Singapore. Given the wide disciplinary focus of COVID-19 and climate change research, we sought to select a range of experts across disciplines. We recruited one expert from each of these disciplines: Public policy, medicine, computational social science. Selecting experts from a range of fields allows results to be contextualized to fields where COVID-19 and climate change research are concentrated, allowing for findings to be drawn on by stakeholders in a range of fields. Based on their expertise, context experts separately developed lists of Telegram groups most relevant to COVID-19 or climate change in Singapore. Groups were chosen based on the number of users in the group, how long the group had been active, and group activity. Each expert developed a list of ten groups independently, and we selected only Telegram groups common to the two groups of experts' lists, within COVID-19 and climate change. The COVID-19 groups selected were: CovidSurvivors, Healing the Divide Discussion, LifeOfUnvaxxed\_Discussion, SG COVID-19 Recovery Support Group, SG Defense against Vaxx, and SG Corona Freedom Lounge. There was a single climate change group: The Other Side of Climate Change. Using tools within Telegram, we downloaded all chats in the groups since inception. Data from the COVID-19 groups was merged into a single file. We analyzed the COVID-19 and climate change data separately (See Table \ref{tab:table_1}). The COVID-19 dataset had 201,438 rows (messages). The climate change dataset contained 2,394 rows. 

\textbf{Data processing}
Many users in the Telegram groups shared links, and we sought to add the text from these links to our dataset (Newspaper3k). Some links were directed to news websites like Channel NewsAsia, The Straits Times, ABC News, New York Post, Seattle Times, Today Online, and Yahoo Finance. The links also included government websites, both Singapore-based and otherwise, and to other telegram groups such as SG Fighters, Village Hotel Sentosa, SGQuarantineOrder, SGVAXInjury, and sgFightScam. Some links could not be scraped due to permission restrictions, and others did not contain any text [COVID-19: 28013/57369 (49\%) links scraped; climate change: 945/1598 (59\%)]. To prepare the texts for model training, we tokenized, lemmatized, and normalized the corpus. See Table \ref{tab:table_2} for an overview of the corpus.


\begin{table}[!htbp] \centering 
\begin{tabular}{ |p{3cm}|p{2cm}|p{2cm}|p{2.5cm}|p{2cm}|p{2cm}|p{1cm}| }
 \hline
 Telegram topics & Rows & Tokens & Normalized tokens & Links present & Successfully scraped links & Users\\
 \hline 
 COVID-19 & 201,438 & 450,667,326 & 21,599,409 & 57,369 & 28,013 & 5152\\
  \hline 
 Climate Change & 2,394 & 30,2853,025 & 13,462,608 & 1,598 & 945 & 70\\
 \hline 
\end{tabular}
\caption{Data overview}
\label{tab:table_2}
\end{table}

\textbf{Feature engineering}
We extracted two broad groups of features: 1) Behavioral (e.g., N of times links shared, N of replies sent), and 2) text (e.g., ngrams, word embeddings). Table \ref{tab:table_3} denotes the behavioral features extracted. 

\begin{table}[!htbp] \centering 
\begin{tabular}{ |p{3cm}|p{6cm}| }
 \hline
 Feature  & Description\\
 \hline 
Messages  & N of times a user sent a message in the chat\\
 \hline 
Words &  N of words used\\
 \hline 
Function words & N of function words\\
 \hline 
Links  & N of links sent by user\\
 \hline 
 Percent links &   \% of posts by a user that were links\\
 \hline 
\end{tabular}
\caption{Behavioural features}
\label{tab:table_3}
\end{table}

We extracted text features as per Table \ref{tab:table_4}. We extracted 71 Linguistic Inquiry and Word Count (LIWC)-based linguistic categories (e.g. friend, gender).  The LIWC software is a computerized way to analyze the word use within a text \cite{boyd2022development}. LIWC calculates the percentage of usage of sets of words that define 80 different linguistic categories, generating an output measure for each of these categories. 

\begin{table}[!htbp] \centering 
\begin{tabular}{ |p{3cm}|p{4cm}| }
 \hline
Name &  Description\\
 \hline 
 Topics &Median topic based on LDA\\
 \hline 
 Sentiment & Median sentiment value\\
 \hline 
 Hate speech & Median hate speech value\\
 \hline 
 Similarity to COVID & Cosine similarity to 'covid'\\
 \hline 
 Similarity to climate & Cosine similarity to 'climate'\\
 \hline 
 2-grams & Most common 2-gram\\
 \hline 
 3-grams & Most common 3-gram\\
 \hline 
 4-grams & Most common 4-gram\\
  \hline 
 LIWC features & LIWC-based features\\
 \hline 
\end{tabular}
\caption{Text features}
\label{tab:table_4}
\end{table}

\textbf{Developing archetypes}
We used principal component analysis (PCA) to create two features (scikit-learn). We then used K-means clustering (scikit-learn), with the PCA features, to separately cluster the COVID-19 and climate change data by users. To choose the optimal number of clusters for each topic, we used the elbow method, based on the within-cluster sum of squares (WCSS) value. We did not include clusters with fewer than 10 users. Two content experts reviewed the text produced by users in each cluster, along with the mean and standard deviation of the features. Content experts proposed archetypes for each cluster independently. Experts then reviewed the archetypes and decided on final archetypes for each cluster. Disagreements were resolved by a third expert. 

\textbf{Classification}
We performed a multi-label classification using a decision tree classifier with the cluster features (scikit-learn). We sought to classify users in the COVID-19 and climate change groups into their respective clusters, with an 80/20 train-test split. The labels for the models were the clusters extracted from the k-means clustering. 

\section{Results}

\begin{table}[!htbp] \centering 
\begin{tabular}{ |p{2cm}|p{2cm}|p{2cm}|p{2cm}|p{2cm}|p{2cm}|}
 \hline
 & \multicolumn{2}{c|}{Climate Change} & \multicolumn{3}{c|}{COVID-19} \\ 
 \hline
 Feature & Strategic Disruptor & Empirical Enthusiast & Inquisitive Moderate & Critical Examiner & Conspiratorial Amplifier \\
 \hline 
 Percent links & 35.5 & 10.3 & 0 & 22 & 49 \\
 \hline 
Function words & 4255 & 28668 & 28 & 109533 & 36555 \\
 \hline 
Messages & 17 & 61.5 & 4 & 423 & 204 \\
 \hline 
Words & 10081 & 72649 & 70 & 263299 & 90084 \\
 \hline 
Links & 4 & 6 & 0 & 1 & 1 \\
 \hline 
Users & 57 & 12 & 4941 & 32 & 175 \\
 \hline 
\end{tabular}
\caption{Median cluster features}
\label{tab:table_5}
\end{table}

We provide median features for the clusters in Table \ref{tab:table_5}. Clusters were significantly different. For example, within the climate change clusters, there was a noted difference in median percent links shared, 35.5\% vs 10.3\%. Similarly, across the climate change clusters, there was much variation in median number of messages (4 vs 423 vs 204). The k-means clustering resulted in two clusters for climate change and three clusters for COVID-19. The climate change data resulted in the following clusters: Strategic Disruptor; Empirical Enthusiast. 

Strategic Disruptor: These users are characterized by active engagement with the intent to challenge mainstream views, often sharing controversial or disruptive content. The texts from this cluster often questioned the integrity of mainstream media and government narratives, suggesting that these entities served corporate interests rather than the public. Common themes included criticism of "green energy" initiatives as profit-driven rather than environmentally motivated and allegations of censorship by major tech companies. Examples: The world has woken up to the lies and corruption of the elites; One of the first glaring lies about the climate agenda...

Empirical Enthusiast: Users in this cluster actively share data and empirical evidence to support their views on climate change, often engaging in detailed discussions. This cluster's texts reflected a blend of information dissemination and opinion sharing. Examples: According to the WEF, why we need to give insects a chance in our diet; Chocolate-covered almonds and almond MnMs certainly aren’t healthier.

The COVID-19 data resulted in the following clusters: Inquisitive Moderate; Critical Examiner; Conspiratorial Amplifier.
Inquisitive Moderate: This cluster was characterized by moderately engaged users who seek clarifications and share content without a strong bias. The median number of messages was four, perhaps indicating limited participation in discussions. Users in this cluster may indicate interest more than actively contribute, occasionally sharing content but not frequently engaging in in-depth discussions. Examples: Can someone explain what this means for us? I found this article interesting, thought I'd share it here. 

Critical Examiner: These highly active users posted content that may be heavily data-driven and analytical. Users in this cluster focused on demographic data, technological impacts, and societal trends. The discussions often included statistical information, and were characterized by a logical and methodical approach to issues such as fertility rates, elder care, and the impact of automation on jobs. Examples: Singapore’s total fertility rate has fallen from 1.82 births per woman in 1990 to 1.14 births in 2020, indicating significant demographic shifts; Voice actors are losing their jobs to AI robots that can mimic human speech with startling accuracy.

Conspiratorial Amplifier: Highly active, possibly amplifying conspiracy theories, and interrogating the integrity of mainstream narratives. Examples: Big Pharma and their government cronies have no interest in our health, it's all about the money; How can we trust the government's narrative when they have been caught lying time and again? Both COVID-19 and climate change clusters demonstrated distinct user engagement patterns and thematic foci. A notable comparison between the clusters for both topics is the presence and tone of criticism. In climate change discussions, Strategic Disruptors express criticism by questioning the integrity and motives of mainstream environmental narratives, focusing on criticisms of green energy initiatives and media portrayals. Their distrust is directed towards the perceived economic and political interests behind environmental policies. In contrast, suspicion in COVID-19 discussions, as seen in the Conspiratorial Amplifiers cluster, is broader and more intense. These users challenge the credibility of public health narratives, government policies, and media reports, often promoting alternative theories and distrustful of official sources. This reflects more profound and wide-ranging criticism compared to the climate change context.

\begin{table}[!htbp] \centering 
\begin{tabular}{ |p{2cm}|p{2cm}|p{2cm}|p{2cm}|p{2cm}|p{2cm}| }
 \hline
 Model & Accuracy & Recall & Precision & F1-Score \\
 \hline 
 COVID-19 & 0.93 & 0.93 & 0.95 & 0.93\\
 \hline 
 Climate & 0.99 & 0.99 & 0.99 &0.99 \\
 \hline 
\end{tabular}
\caption{Model metrics}
\label{tab:table_6}
\end{table}

We presented classification metrics for the exploratory classification model in Table \ref{tab:table_6}. The model performed well at classifying users into clusters we previously developed. The classification model enhances our ability to identify and understand these archetypes, offering valuable insights for future efforts to combat misinformation and enhance public discourse on pressing global issues.

\section{Discussion}
\textbf{Implications of Findings} 
The clustering analysis reveals a complex ecosystem of user archetypes within Telegram groups discussing climate change and COVID-19. Understanding these archetypes is crucial for developing targeted interventions to mitigate misinformation and enhance public discourse. Although this project focuses on clustering and classification models, these insights could inform future empirical investigations aimed at designing specific interventions \cite{goyal2023predicting,goyal2024using}.


\textbf{Limitations}
The data is sourced from specific Telegram groups, which may not represent the broader population's views or behaviors. The clustering method used may not capture the full complexity of user behaviors and motivations, as it relies on text and behavioral features that may not encompass all relevant factors. Additionally, while the classification model demonstrated high accuracy, recall, precision, and F1-scores, it is based on the initial clusters and may require further validation with new data to ensure its robustness and generalizability. Finally, the focus on COVID-19 and climate change discussions means that the findings may not be directly applicable to other topics or contexts, necessitating further research to explore user archetypes in different thematic areas.


\bibliographystyle{ACM-Reference-Format}
\bibliography{sample-base}


\end{document}